# Radio Astronomy Data Transfer and eVLBI using KAREN

*Stuart Weston[1], Timothy Natusch[1] and Sergei Gulyaev[1]*

[1]Institute for Radio Astronomy and Space Research, Auckland University of Technology,
Private Bag 92006, Auckland 1142, New Zealand
stuart.weston@aut.ac.nz, tim.natusch@aut.ac.nz, sergei.gulyaev@aut.ac.nz

## Abstract

Kiwi Advanced Research and Education Network (KAREN) has been used to transfer large volumes of radio astronomical data between the Radio Astronomical Observatory at Warkworth, New Zealand and various international organizations involved in joint projects and VLBI observations. Here we report on the current status of connectivity and on the results of testing different data transfer protocols. We investigate new UDP protocols such as 'tsunami' and UDT and demonstrate that the UDT protocol is more efficient than 'tsunami' and 'ftp'. We also report on the tests on direct data streaming from the radio telescope receiving system to the correlation centre without intermediate buffering or recording (real-time eVLBI).

## 1. Introduction

The Warkworth Radio Astronomical Observatory (WRAO) is located some 60 km north of the city of Auckland, near the township of Warkworth. The observatory is operated by the Institute for Radio Astronomy and Space Research (IRASR) of the Auckland University of Technology (AUT). The observatory's 12-m radio telescope operates in three frequency bands centered around 1.4, 2.4 and 8.4 GHz. This fast-slewing (5° per second in Azimuth) antenna is well suited to the purposes of geodetic VLBI (Very Long Baseline Interferometry), for spacecraft navigation and tracking. It is also effectively used for astronomical VLBI research in conjunction with large radio telescopes [1].

Each VLBI session may result in many Terabytes of data recorded, so transfer of data to a data processing (correlation) centre is an issue of great importance for VLBI. Previously data was recorded to magnetic tapes, these were then sent via traditional means such as post/courier to the correlation centre. More recently with the reduced cost and increased capacity of hard disk storage, the data has been recorded to removable disk arrays which can then also be sent physically to the correlation centre. Recently with the advent of eVLBI the data is sent via high speed networks to the correlation centers in real-time.

The IRASR collaborates with a number of international partners. This research collaboration can be broken into three major research topics/groups:

- Astrophysical VLBI and eVLBI observations in the framework of the Australian Long Baseline Array (LBA) for the study of physics and origin of extragalactic and galactic radio sources: active galactic nuclei, radio galaxies, supernova remnants and star formation regions. These observations require data to be sent from the 12-m radio telescope to CSIRO (Sydney) and/or to the Curtin University, Perth for correlation and imaging [2].

- Observation and navigation of inter-planetary spacecraft, as well as ground tracking services for a variety of space missions. In 2010 the 12-m radio telescope participated in VLBI observations of JAXA's IKAROS and Akatsuki spacecraft; ESA's Mars Express was successfully detected. Data was transferred electronically (e-transfer) directly to Metsähovi, Finland and to the Joint Institute for VLBI in Europe (JIVE) [3].

- Geodetic VLBI and regular IVS (International VLBI Service for Geodesy and Astrometry) observations of a large group of quasars, which uniquely support the International Celestial Reference Frame [4]. The observational data are to be sent to data correlation centres located in the United States Naval Observatory (USNO) and the Max Planck Institute for Radio Astronomy (MPIfR) in Bonn (Germany).

With the connection of the WRAO to KAREN network, our intention is to optimize the use of KAREN for transferring large volumes of observational data to our partners in Australia, Asia, North America and Europe and for conducting real-time eVLBI. In this paper we discuss the use of FTP over TCP/IP protocol for transferring data, and compare the performance of new protocols which are being used in radio astronomy such as 'tsunami' and UDT (UDP-based Data Transfer) via the network protocol UDP.

## 2. Network Protocols and Connectivity Status

Point to point with no hops FTP is efficient, but as the number of hops in the route increases and the incidence of lost packets and collisions increases the TCP congestion avoidance algorithm becomes a severe limitation to the throughput that can be achieved.

'Tsunami' is an UDP file transfer protocol developed by Jan Wagner of the Metshovi Radio Observatory in 2007 [5]. This is the protocol of choice for sending files to Bonn for the IVS observations, as stipulated by the MPIfR. Another UDP protocol called UDT was developed in the University of Illinois in 2005 [6, 7]. UDT was investigated by our Australian partners (private communication Chris Phillips, CSIRO) in 2008-09. This now appears to have matured and is of further interest and warrants further investigation.

Table 1 presents the destinations that connectivity has been achieved with (column 1), the protocols that have been verified for data transfer (columns 2 and 3) and command line access to remote servers for initiating data transfers (column 4).

Table 1: Connectivity established between the IRASR and its VLBI partners via KAREN

| Destination | Protocol | | Command | Date |
|---|---|---|---|---|
| CSIRO (Australia) | UDP | Tsunami, UDT | ssh | 01/04/2010 |
| Bonn (Germany) | UDP | Tsunami, UDT | ssh | 01/06/2010 |
| JIVE (Netherlands) | UDP | - | iperf | 27/07/2010 |
| Metsähovi (Finland) | UDP | Tsunami, UDT | ssh | 21/07/2010 |
| USNO (United States) | UDP | - | ssh, iperf | 15/01/2011 |
| GSI (Japan) | UDP | Tsunami, UDT | ssh, iperf | 10/1/02011 |

## 3. Data Transfer Tests and Results

Table 2 presents the results of data transfer tests between the IRASR and data processing centers in Bonn and Metsähovi. The results were obtained by transferring an actual 16 bit VLBI file produced in observations with the 12-m radio telescope. Column 2 shows the protocol used, column 3 gives the amount of data sent in bytes, column 4 provides the time it took to transfer the data and column 5 shows an average throughput rate over the period. The data was transferred from the IRASR's IBM Blade server via the KAREN network using the default settings for each protocol with no tunning. The light path between New Zealand and Europe is shown in Figure 1.

Table 2: Data Transfer results: IRASR to Bonn and Metsähovi

| Route | Protocol | Bytes | Time (s) | Throughput (Mbps) |
|---|---|---|---|---|
| AUT – Bonn | Ftp | 65G | 8016 | 65 |
| AUT – Metsähovi | Ftp | 3.1G | 432 | 61 |
| AUT – Bonn | Tsunami | 65G | 3466 | 151 |
| AUT – Metsähovi | Tsunami | 65G | 4979 | 105 |
| AUT – Bonn | UDT | 65G | 1920 | 273 |
| AUT – Metsähovi | UDT | 65G | 1157 | 453 |

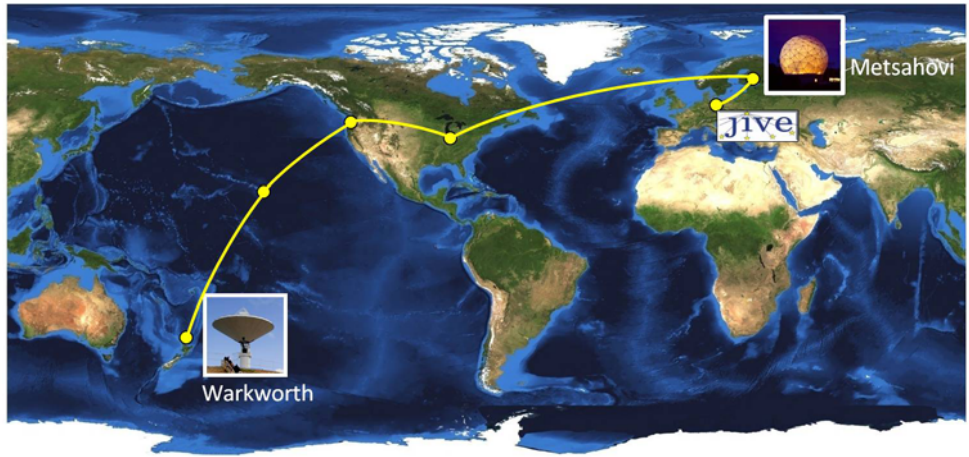

Figure 1.  The lightpath New Zealand – Metsähovi – JIVE. Satellite image: Blue Marble Next Generation. Courtesy: NASA Visible Earth

The main purpose of these tests was to compare the performance of different data transfer protocols. Data in Table 2 clearly demonstrate the advantage of the UDT protocol. It is up to 4 times faster than 'tsunami' protocol and almost one order of magnitude faster than the standard 'ftp' protocol. Tests were conducted repeatedly over several days and at different times resulting in slightly different average rates without changing the main conclusion that the UDT protocol is superior to 'ftp' and 'tsunami' . The difference in data transfer rates for different end points can be explained by different number of hops and different performance of sections along the routes. A traceroute command from the AUT Blade to the IP address in Metsähovi gave a route of 14 hops. Re-issuing this command repeatedly over several months showed that the route appears stable, without any changes. A high number of hops on the route (14) demonstrates the complexity of the path and explains why data transfers via protocols such as FTP are not efficient.

A series of additional data transfer sessions was conducted in 2010 and 2011.

In August 2010 observations were conducted of the ESA's Mars Express spacecraft orbiting Mars. This data was transmitted to Metsähovi via KAREN. The data of 24 August which totaled 86 GB was sent using the UDP protocol 'tsunami' immediately after the observational session. The next set of observational data (26 August) totaling 187 Gbytes was sent via UDT. Average rates sustained were 250 and 300 Mbps respectively.

Another set of experiments was conducted in September 2010 aiming to test the 'tsunami' protocol for streaming VLBI data directly from the radio telescope receiving/recording system (PCEVN) via KAREN to Metsähovi. This test was an important step towards real-time eVLBI. Initial tests from the WRAO to the IBM Blade server at the AUT City Campus showed that the 'tsunami' rates of 485 Mbps with no lost packets was sustainable. However, when streaming data from Warkworth to Metsähovi many thousands of lost packets occurred and a sustainable rate of 350 Mbps was achieved. This is significantly lower than the rate of 512Mbps required for the real-time eVLBI streaming of 8bit data to Metsähovi.

In June 2010 file transfer tests to the correlator site at Curtin University, Perth were conducted. Using 'tsunami', rate of 300 Mbps was achieved, while UDT was superior with 400 Mbps. In December 2010 the first eVLBI tests from the Mk5B recorder at the Observatory to the server at CSIRO in Australia using UDP were conducted. The required data rate of 512Mbps was achieved sustainably. In January 2011 eVLBI tests with CSIRO where undertaken to stream data directly from the Mk5B to the correlator: a rate of 512Mbps was achieved for 8 channels at 2 bit resolution. In February 2011 successful real-time eVLBI tests and demonstration were undertaken between the New Zealand 12-m radio telescope and the LBA antennas in Australia [8].

In addition, in January 2011 connectivity was established with the Geographical Survey Institute of Japan (GSI) and tests were conducted between the blade server at the AUT City Campus, Auckland and the GSI using "iperf". Rate of 512 Mbps was achieved with some packet loss (0.00021%), but at slower rates there was no packet loss. When using UDT for transferring files, the loss of packets at a low rate would not be a major factor. Currently the KAREN link between New Zealand and the USA (LAX) is relatively lightly loaded with the maximum capacity of 1 Gbps. The tests where conducted to transfer a 16 GB Mk5B vector file from the WRAO. A similar rate of about 350 Mbps was achieved with both 'tsunami' and UDT protocols.

The next step is to test the more efficient UDT protocol for direct data streaming. However, to make use of UDT instead of 'tsunami' for streaming requires the VLBI software to be modified. Some programming effort is needed as the network code is merged within the application code rather than being maintained in a separate library.

## 4. Conclusion

It was demonstrated that the use of the UDP protocol for radio astronomical data transfer has the required sustained data transfer rates for eVLBI, but UDT has some advantages over 'tsunami':

- UDT is a better citizen on the network leaving bandwidth for TCP and other UDP protocols, which is very important on a shared network such as KAREN.

- UDT has an application programming interface (API) allowing easy integration with existing or future applications.

We investigated the ability to stream collected data via UDT and modified the Curtin 16 bit to 2 bit conversion program for data streaming to a remote server in a fast and efficient manner ready for the correlator. This has been successfully demonstrated between Warkworth and the IBM Blade Server at AUT. Currently the JIVE Mk5 code is being modified to stream data via UDT to a real-time digital autocorrelation spectrometer implemented on IBM InfoSphere Streams [9].

We have found KAREN to be a very useful tool for transmitting data to our international partners, and the IRASR will be extending its use over the coming months. Of future benefit to our work to stream data real-time to the international correlation centers would be the ability to reserve bandwidth as a logical pipe within the KAREN bandwidth for the duration of an experiment.

## Acknowledgements

This research was supported by the Ministry for Economic Development's SKA Grant. Warkworth connectivity via KAREN was made possible by the Remote Site Connectivity Fund provided by the Ministry of Research Science and Technology (MoRST) on behalf of the New Zealand Government. We are grateful to AUT and KAREN staff, as well as to our VLBI partners in Australia, Japan, Finland, USA, Germany and the Netherlands for their support and assistance.